\documentclass[a4paper, 12pt]{article}
\RequirePackage{amsthm,amsmath,amsfonts,amssymb}
\RequirePackage[authoryear]{natbib}

\usepackage{amsmath, bbm}
\usepackage{graphicx}
\usepackage{enumerate}

\usepackage{url} 
\usepackage{comment}
\usepackage{subcaption}
\usepackage{amsfonts} 
\usepackage{color}
\usepackage{mathtools}

\usepackage{amsmath, amssymb}
\usepackage{graphicx, epsfig}
\usepackage{epstopdf}
\usepackage{enumerate}
\usepackage{natbib}
\usepackage{url, xcolor}

\newtheorem{theorem}{Theorem}[section]
\newtheorem{lemma}{Lemma}[section]

\newcommand{\ve}[1]{{\mbox{\boldmath ${#1}$}}}

\begin{document}

\def\spacingset#1{\renewcommand{\baselinestretch}%
{#1}\small\normalsize} \spacingset{1}
\title{\bf Direct estimation and inference of higher-level
correlations from lower-level measurements with
applications in gene-pathway and proteomics
studies}
  \author {
    \small
     Yue Wang  \\
       \small
   Department of Biostatistics and Informatics, University of Colorado Anschutz Medical Campus\\
    \small
    and\\
    \small
    Haoran Shi\\
       \small
     School of Mathematical and Statistical Sciences, Arizona State University \\
    } 
  \maketitle

\begin{abstract}
{This paper tackles the challenge of estimating correlations between higher-level biological variables (e.g., proteins and gene pathways) 
when only lower-level measurements are directly observed (e.g., peptides and individual genes). 
Existing methods typically aggregate lower-level data into higher-level variables and then estimate correlations based on the aggregated data. However, different data aggregation methods can yield varying correlation estimates as they target different higher-level quantities. Our solution is a latent factor model that directly estimates these higher-level correlations from lower-level data without the need for data aggregation.
We further introduce a shrinkage estimator to ensure the positive definiteness and improve the accuracy of the estimated correlation matrix.  Furthermore, we establish the asymptotic normality of our estimator, enabling efficient computation of $p$-values for the identification of significant correlations. The effectiveness of our approach is demonstrated through comprehensive simulations and the analysis of proteomics and gene expression datasets. 
We develop the {\tt R} package {\tt highcor} for implementing our method.
}
\end{abstract}
Correlation matrix, latent variables, shared variables, shrinkage estimation
\section{Introduction}\label{sec:intro}


Data aggregation from a lower level of granularity to a higher level is common in biological research. 
For example, in proteomics studies, 
protein abundances are { often} derived by aggregating peptide-level information \citep{silva2006absolute}. 
In gene pathway analyses, individual gene expression profiles { can be} aggregated to quantify the expression level of each gene pathway \citep{tomfohr2005pathway, edelman2006analysis, lee2008inferring}. 
These aggregated datasets are crucial for downstream statistical analyses aimed at uncovering mechanisms in various higher-level biological processes.
However, the reliability of these analyses is often contingent on the data aggregation approach, particularly in how shared elements are handled. For example, in proteomics, multiple proteins may share peptides, while in gene-pathway analysis, different pathways might include the same genes. 
Proposing new and statistically supported methods to incorporate shared variables into downstream statistical analyses can benefit many fields of omics biology.


This paper addresses the challenge of estimating correlations between high-level biological variables, such as proteins and gene pathways, 
when only lower-level measurements, such as peptides and individual genes, are directly observed.
These higher-level correlations, such as protein-protein interactions and pathway co-expression, represent higher-level systemic functions of the cell and the organism \citep{pita2018pathway, rao2014protein}. 
Most existing methods first aggregate lower-level data into higher-level variables and then perform correlation estimation based on the aggregated data.
These aggregation methods can be divided into two categories: methods that utilize selected variables (e.g., unique or top variables) and those that incorporate all variables.
Selected variables-based methods are often based on the sum or mean of the selected variables. 
In proteomics studies, such methods include the sum or mean of all unique peptides and the sum of the 3 most intense unique peptides \citep{silva2006absolute, malmstrom2009proteome}. In gene-pathway analyses, such methods include the mean of condition-responsive genes \citep{lee2008inferring} and the mean of the top 50\% genes \citep{hwang2012comparison}. 
Methods that use all variables are more diverse, including the use of the sample mean \citep{levine2006pathway}, the Tukey median polish \citep{tukey1977exploratory, choi2014msstats}, the principal component analysis \citep{tomfohr2005pathway}, 
and regression models (e.g., SCAMPI, \citealp{gerster2014statistical}). 
While these methods are widely applied, the effects of different aggregation techniques on the accuracy of higher-level correlation estimation remain ambiguous. 
Other related methods in the field of gene pathway analysis include unsupervised, single-sample enrichment methods, such as ssGSEA (single-sample gene set enrichment analysis, \citealp{barbie2009systematic}), PLAGE (pathway-level analysis of gene expression, \citealp{tomfohr2005pathway}), and GSVA (gene set variation analysis, \citealp{hanzelmann2013gsva}). These methods compute an enrichment score for each gene pathway and each individual sample based on various data transformations, such as the standardization used by PLAGE or the empirical cumulative distribution methods used by ssGSEA and GSVA. As such, the resulting enrichment scores are not typically used for estimating correlations between gene pathways, as these data transformations disrupt the correlation structures present in the original data. 
We will present numerical analyses to demonstrate that all these existing methods may lead to sub-optimal estimates of higher-level correlations.

 We develop an innovative latent factor model to address the challenge of estimating higher-level correlations without relying on data aggregation. 
 Our method conceptualizes higher-level variables as latent variables, which are linked to lower-level variables via a binding matrix derived from domain knowledge. Drawing inspiration from real biological studies, we assume that each higher-level variable is associated with at least two unique lower-level variables to ensure model identifiability. 
 Based on this assumption, we introduce a direct estimator formulated through a novel estimating equation that connects higher-level correlations with those at the lower level. 
To enhance estimation accuracy, we introduce a class of shrinkage estimators. By choosing an optimal weight via cross-validation, the shrinkage estimator is shown to achieve superior accuracy. We also establish the asymptotic normality of the direct estimator, facilitating the efficient calculation of $p$-values for identifying significant higher-level correlations.
The efficacy of our proposed method is validated through extensive simulations. We demonstrate its effectiveness by analyzing a proteomics study and a gene expression dataset. 


Throughout the paper, we use normal typeface to denote scalars, bold lowercase typeface to denote vectors, and uppercase typeface to denote matrices.  For any vector ${\bf v} \in \mathbb{R}^p$, we use $v_j$ to denote the $j$-{th} element of $\bf v$ for $j = 1, \ldots, p$. For any matrix ${M} \in \mathbb{R}^{n \times p}$, let $ {\bf m}_j$ denote the  $j$-{th} column of $M$ for $j = 1, \ldots, p$.
We use $m_{ij}$ or $(M)_{ij}$ to denote the $(i,j)$ entry of $M$ for $i = 1, \ldots, n$ and $j = 1, \ldots, p$. 
For any index set $\mathcal{I} \subset \{1, \ldots, p\}$, let ${\bf v}_\mathcal{I}$ and ${M}_{\mathcal{I}}$ denote the subvector of ${\bf v}$ whose elements are indexed by $\mathcal{I}$ and the submatrix of ${M}$ whose columns are indexed by $\mathcal{I}$, respectively. 
Let $I(\mathcal{A})$ denote the indicator function of the event $\mathcal{A}$; i.e., $I(\mathcal{A}) = 1$ if $\mathcal{A}$ is true, and $I(\mathcal{A}) = 0$ otherwise. 
 We denote
$ \|{\bf v}\|_0 = \sum_{j = 1}^p I(v_j \neq 0), ~~ \|{\bf v}\|_q = \left(\sum_{j = 1}^p |v_j|^q\right)^{1/q} $ for any $ 0 < q < \infty$, $\|{\bf v}\|_{\infty} = \max_j|v_j|  $, 
 $\|{ M}\|_q = \sup_{\|{\bf v}\|_q = 1}\|M{\bf v}\|_q$ for any $q > 0$ and $\|{M}\|_F^2 = \sum_{i=1}^n \sum_{j=1}^p m_{ij}^2$. 
 Let $I_n$ and ${\bf 1}_n$ denote the $n \times n$ identity matrix and the $n \times 1$ vector with all ones, respectively. 
 Let $\mbox{vec}(M)$ denote the vector stacking the columns of $M$ on top of one another:
 \(
 \mbox{vec}(M) = (m_{1,1}, \ldots, m_{n,1}, m_{1,2}, \ldots, m_{n,2}, \ldots, m_{1,p}, \ldots, m_{n,p})^\intercal.
 \)
 For two matrices $M_1$ and $M_2$, let $M_1 \otimes M_2$ denote the Kronecker product of $M_1$ and $M_2$. 
  Finally, for any square matrix ${S}$, we denote 
 the sum of the diagonal elements of $S$ as $\mbox{tr}({S})$.


\section{Model}\label{sec:method}


We consider a data set with $n$ subjects that measures $q$ lower-level variables, denoted ${\bf z}_i = (z_{i1}, \ldots, z_{iq})^\intercal$ for $i = 1, \ldots, n$.
These variables map onto $p$ {\it latent} higher-level variables, denoted ${\bf x}_i = (x_{i1}, \ldots, x_{ip})^\intercal$, through a binding matrix $A \in \mathbb{R}^{q \times p}$. The mapping matrix $A$, obtained from domain knowledge, has 0-1 entries, where $a_{kj} = 1$ if the $k$-th lower-level variable belongs to the $j$-th higher-level variable, and $a_{kj} = 0$ otherwise. 
{Note that each lower-level variable may belong to one or more higher-level variables. 
We call those lower-level variables belonging to only one higher-level variable {\it unique variables}, 
and those belonging to multiple higher-level variables are called {\it shared variables}.
}
In this paper, we focus on the estimation and inference of the higher-level correlations. 
More specifically, assuming $\mathbb{E}[{\bf x}_i] = {\bf 0}$ and $\mbox{cov}({\bf x}_i) = \Sigma = (\sigma_{lk})_{l,k = 1, \ldots, p}$, 
our goal is to estimate
\(
R = \text{diag}(\Sigma)^{-1/2}\Sigma\text{diag}(\Sigma)^{-1/2},
\)
where $\text{diag}(\Sigma) = \text{diag}(\sigma_{11}, \ldots, \sigma_{pp})$.
Accomplishing this goal can help elucidate the complex relationships between various higher-level biological processes. 


If the higher-level variables (i.e., ${\bf x}_i$'s) were directly observed, a natural estimator of $R$ is the sample correlation matrix $\sum_{i=1}^n {\bf x}_i {\bf x}_i^\intercal /n$.  
However, when only lower-level variables (${\bf z}_i$) are directly measured, calculating the sample correlation matrix in this way requires estimating 
${\bf x}_i$ by aggregating the lower-level variables ${\bf z}_i$ according to the binding matrix $A$.
As discussed in the Introduction, existing aggregation methods apply a mathematical operator (e.g., sum, mean, or median) or model (e.g., linear model) to all or a pre-defined set of variables (e.g., unique variables). 
However, different aggregation methods may lead to different estimates of higher-level correlations which may have varying levels of bias. We refer readers to the supplementary document for the theoretical analyses of selected aggregation methods. Numerical evaluations of these aggregation methods are provided in Sections \ref{sec:simu} and \ref{sec:realdata}, as well as in the supplementary document.

 We propose a novel approach that directly estimates $R$ based on the lower-level data ${\bf z}_i$'s without inferring each ${\bf x}_i$. 
Motivated by the binding relationship characterized by $A$, our approach is based on the following latent factor model: 
\begin{align}\label{071020:1}
          {\bf z}_i = A{\bf x}_i + \ve \varepsilon_i,
    \end{align}
    where 
    $\ve \varepsilon_i$ is assumed independent of ${\bf x}_i$ with $\mathbb{E}(\ve \varepsilon_i) = {\bf 0}$ and $\mbox{cov}(\ve \varepsilon_i) = \Gamma = \text{diag}(\gamma_1, \ldots, \gamma_q)$, representing various technical noises and measurement errors. 
    We do {\it not} assume any distributional assumptions (such as normality) about the data.

Unfortunately, model (\ref{071020:1}) does not guarantee the identifiability of $\Sigma$ or $R$ without additional conditions. 
To see this, we note from (\ref{071020:1}) that 
\(
C = A\Sigma A^\intercal + \Gamma,
\)
where $C = \mbox{cov}({\bf z}_i)$. 
Suppose the binding matrix $A$ has rank $p$ and $\mbox{Cov}(\Gamma) = \sigma^2 I_q$ for some $\sigma^2 > 0$. 
In this case, $A^\intercal A$ is invertible, and we can obtain
\(
    (A^\intercal A)^{-1} A^\intercal C A(A^\intercal A)^{-1} = \Sigma + \sigma^2 (A^\intercal A)^{-1}, 
\)
implying that 
\(
\Sigma = (A^\intercal A)^{-1}A^\intercal(C - \sigma^2 I_q)A(A^\intercal A)^{-1}. 
\)
Thus, different values of $\sigma^2$ that are less than the smallest eigenvalue of $C$ lead to different values of $\Sigma$, showing that $\Sigma$ is not identifiable without additional conditions. 

The identifiability of model \eqref{071020:1} has been widely discussed in the existing literature on latent factor models. 
Well-known identification results rely on the constraint that the latent factors are mutually uncorrelated \citep{fan2016projected, fan2021robust}. This ``zero-correlation" constraint allows for the use of principal component analysis and its variants to estimate the latent factors. 
This constraint, however, is inappropriate for our problem where the latent higher-level factors have inherent correlations. 
Alternative identification results rely on special structures of the loading/binding matrix $A$. For example, \cite{lam2012factor} assumes that the rows of $A$ are orthonormal, i.e., $A^\intercal A = I_p$; \cite{bai2012statistical} considers a scenario that $A$ can be partitioned into $A_1 \in \mathbb{R}^{p \times p}$ and $A_2 \in \mathbb{R}^{(q-p) \times p}$ with $A_1 = I_p$ and $A_2$ estimated from the data. 
While these constraints have been proved useful in various applications, they do not apply to our problem where $A$ is a pre-determined binary matrix. 

Motivated by the existence of unique variables in real applications, we propose the {\it unique-variable} condition (UVC) on the mapping matrix $A$.
\begin{description}
    \item[UVC] 
     For every $l \in \{1, \ldots, p\}$, there exist at least two indices $j \in \{1, \ldots, q\}$ such that $a_{jl}=1$ and $a_{jk} =0$ for all $k \neq l$.
\end{description}  
UVC postulates the existence of at least {\it two lower-level variables}, which are associated with one and only one higher-level variable. 
Under UVC, we establish the identifiability of the covariance matrix $\Sigma$ in the following result.
\begin{lemma}\label{thm1}
Suppose that UVC is satisfied. Then, under model (\ref{071020:1}), the covariance matrix $\Sigma$ can be uniquely determined from $C$ and $A$ by
\begin{align}\label{thm1:eq1}
\sigma_{ll}= \frac{1}{|\mathcal{S}_l|(|\mathcal{S}_l| - 1)} \sum_{i,j \in \mathcal{S}_l; i\neq j} c_{ij} ~\mbox{and}~
\sigma_{lk}= \frac{1}{|\mathcal{S}_l||\mathcal{S}_k|} \sum_{i \in \mathcal{S}_l, j \in \mathcal{S}_k} c_{ij},
\end{align}
  for $l, k \in \{1, \ldots, p\}$ and $l \neq k$, where $\mathcal{S}_l \subset \{1, \ldots, q\}$ denotes the set of indices of the unique lower-level variables belonging to the $l$-th higher-level variable for $l = 1, \ldots, p$.
\end{lemma}
The proof of Lemma \ref{thm1} is given in the supplementary document.
This result not only guarantees the identifiability of $\Sigma$ but also provides a set of estimating equations that enable direct estimation of higher-level correlations from lower-level data. We will discuss this direct estimation approach and its theoretical properties in the section below. 
\section{Direct Estimation and Inference}\label{direct:est}

In this section, we discuss the direct procedure for the estimation and inference of $R$.  
Letting
$\widehat{c}_{jk} = (n-1)^{-1}\sum_{i=1}^n z_{ij}z_{ik}$, 
\eqref{thm1:eq1} leads to the direct estimator 
\(
\widehat{R}_{\text{dir}} = \text{diag}(\widehat{\Sigma}_{\text{dir}})^{-1/2}\widehat{\Sigma}_{\text{dir}}\text{diag}(\widehat{\Sigma}_{\text{dir}})^{-1/2},
\)
where
$\widehat{\Sigma}_{\text{dir}} = \left(\widehat{\sigma}_{lk}\right)$ with
\begin{equation}\label{direct:est:1}
\widehat{\sigma}_{ll}= \frac{1}{|\mathcal{S}_l|(|\mathcal{S}_l| - 1)} \sum_{i,j \in \mathcal{S}_l; i\neq j} \widehat{c}_{ij} ~\mbox{and}~
\widehat{\sigma}_{lk}= \frac{1}{|\mathcal{S}_l||\mathcal{S}_k|} \sum_{i \in \mathcal{S}_l, j \in \mathcal{S}_k} \widehat{c}_{ij},
\end{equation}
  for $l, k \in \{1, \ldots, p\}$ and $l \neq k$.
  Since $\widehat{c}_{jk}$ is an unbiased estimator of $c_{jk}$, it is easy to see that $\widehat{\Sigma}_{\text{dir}}$ is an unbiased estimator of $\Sigma$. 
Recalling that $\mathcal{S}_l$ denotes the set of indices of the unique lower-level variables belonging to the $l$-th higher-level variable for $l = 1, \ldots, p$, 
we introduce additional notation:   
\[
\mathcal{I}_{ll} = \left\{ i + (j-1)q: ~\forall i \in \mathcal{S}_l, j \in \mathcal{S}_l, i \neq j \right\} 
\mbox{ and } 
\mathcal{I}_{lk} = \left\{ i + (j-1)q: ~\forall i \in \mathcal{S}_l, j \in \mathcal{S}_k \right\}
\]
for $l \neq k \in \{1, \ldots, q\}$. 
Also, denote ${\bf m}_{l_1l_2} = \left(I(1 \in \mathcal{I}_{l_1l_2}), \ldots, I(q^2 \in \mathcal{I}_{l_1l_2})\right)^\intercal \in \mathbb{R}^{q^2} $ for $l_1, l_2 = 1, \ldots, q$,  and $V = \mathbb{E}[{\bf z}_i{\bf z}_i^\intercal \otimes {\bf z}_i{\bf z}_i^\intercal] - \mbox{vec}(C)\mbox{vec}(C)^\intercal$. 
The following result establishes the asymptotic distribution of $\sqrt{n}\mbox{vec}(\widehat{\Sigma}_{\text{dir}} - \Sigma)$, which serves as the basis for valid inference about $R$.
\begin{lemma}\label{cov:upc:prop}
Suppose all assumptions in Lemma \ref{thm1} hold. 
Consider the direct estimator in (\ref{direct:est:1}). As $n \rightarrow \infty$,  we have 
\(
\sqrt{n}\mbox{vec}(\widehat{\Sigma}_{\text{dir}} - \Sigma) \stackrel{d}{\rightarrow} N_{p^2}(0, \Theta), 
\)
where $\Theta = (\theta_{rs})_{r,s = 1, \ldots, p^2}$ with $r = l_1 + (k_1 - 1)p$ and $s = l_2 + (k_2 - 1)p$, and 
\[
\theta_{rs} = 
\begin{cases}
\frac{1}{|\mathcal{S}_{l_1}| |\mathcal{S}_{l_2}| (|\mathcal{S}_{l_1}| - 1) (|\mathcal{S}_{l_2}| - 1)} {\bf m}_{l_1l_1}^\intercal V {\bf m}_{l_2l_2} & \mbox{ if } l_1 = k_1, l_2 = k_2 \\
\frac{1}{|\mathcal{S}_{l_1}| |\mathcal{S}_{l_2}| (|\mathcal{S}_{l_1}| - 1) |\mathcal{S}_{k_2}|} {\bf m}_{l_1l_1}^\intercal V {\bf m}_{l_2k_2} & \mbox{ if } l_1 = k_1, l_2 \neq k_2 \\
\frac{1}{|\mathcal{S}_{l_1}| |\mathcal{S}_{l_2}| |\mathcal{S}_{k_1}| |\mathcal{S}_{k_2}|} {\bf m}_{l_1k_1}^\intercal V {\bf m}_{l_2k_2} & \mbox{ if } l_1 \neq k_1, l_2 \neq k_2.
\end{cases}
\]
\end{lemma}
Letting \(
{\bf f}_{lk} = \left( \sigma_{ll}^{-1/2}\sigma_{kk}^{-1/2}, - r_{lk} \sigma_{ll}^{-1}/2, -r_{lk} \sigma_{kk}^{-1}/2 \right)^\intercal
\), we have the following result. 
\begin{theorem}\label{thm:cor}
Consider the direct correlation estimator $\widehat{R}_{\text{dir}} = \left(\widehat{r}_{lk}\right)_{j,k = 1, \ldots, p}$. For any $l \neq k$, we have 
\(
\sqrt{n}(\widehat{r}_{lk} - r_{lk}) \stackrel{d}{\rightarrow} N(0, \delta_{lk}^2) \mbox{ as } n \rightarrow \infty, 
\)
with $\delta_{lk}^2 = {\bf f}_{lk}^\intercal \Upsilon_{lk} {\bf f}_{lk}$, where
\[
\Upsilon_{lk} = 
\begin{bmatrix}
\frac{ {\bf m}_{lk}^\intercal V {\bf m}_{lk} }{|\mathcal{S}_l|^2 |\mathcal{S}_k|^2 } &  
\frac{ {\bf m}_{lk}^\intercal V {\bf m}_{ll} }{|\mathcal{S}_l|^2 |\mathcal{S}_k| (|\mathcal{S}_l| - 1) }
& \frac{ {\bf m}_{lk}^\intercal V {\bf m}_{kk} }{|\mathcal{S}_k|^2 |\mathcal{S}_l| (|\mathcal{S}_k| - 1) }
\\
\frac{ {\bf m}_{lk}^\intercal V {\bf m}_{ll} }{|\mathcal{S}_l|^2 |\mathcal{S}_k| (|\mathcal{S}_l| - 1) }
& \frac{ {\bf m}_{ll}^\intercal V {\bf m}_{ll} }{|\mathcal{S}_l|^2 (|\mathcal{S}_l| - 1)^2 } 
& \frac{ {\bf m}_{ll}^\intercal V {\bf m}_{kk} }{|\mathcal{S}_l| |\mathcal{S}_k| (|\mathcal{S}_l| - 1) (|\mathcal{S}_k| - 1) }
\\
\frac{ {\bf m}_{lk}^\intercal V {\bf m}_{kk} }{|\mathcal{S}_k|^2 |\mathcal{S}_l| (|\mathcal{S}_k| - 1) }
& 
\frac{ {\bf m}_{ll}^\intercal V {\bf m}_{kk} }{|\mathcal{S}_l| |\mathcal{S}_k| (|\mathcal{S}_l| - 1) (|\mathcal{S}_k| - 1) }
& 
\frac{ {\bf m}_{kk}^\intercal V {\bf m}_{kk} }{|\mathcal{S}_k|^2 (|\mathcal{S}_k| - 1)^2 }\\
\end{bmatrix}.
\]
\end{theorem}

The proofs for Lemma \ref{cov:upc:prop} and Theorem \ref{thm:cor} are given in the supplementary document. 
In practice, it is of particular interest to detect pairs of higher-level variables whose absolute correlations are beyond a pre-specified magnitude $\xi \geq 0$. 
Specifically,
consider
\begin{equation}\label{hypo:1}
H_0: |r_{lk}| \leq \xi \mbox{ vs. } H_1: |r_{lk}| > \xi. 
\end{equation}
In this case, since the normal distribution satisfies the monotone likelihood ratio property, 
to control the type-I error rate, it suffices to consider the null distribution of $|r_{lk}| = \xi$. This leads to the following pair of test statistics
\begin{equation}\label{test:stat}
T^+_{lk} = \frac{\sqrt{n} \left(\widehat{r}_{lk} - \xi\right)_{+}
}{ \sqrt{\widehat{\delta}_{lk}^2} } \mbox{ and }  T^-_{lk} = \frac{\sqrt{n} \left(\widehat{r}_{lk} + \xi\right)_{-}
}{ \sqrt{\widehat{\delta}_{lk}^2} }, 
\end{equation}
where $a_+ = \max(a, 0), a_- = \min(a, 0)$ for any $a \in \mathbb{R}$, and
$\widehat{\delta}_{lk}^2$ is the plug-in estimate of $\delta_{lk}^2$. 
According to Theorem \ref{thm:cor}, 
an asymptotically valid  two-sided $p$-value is 
\begin{equation}\label{test:stat2}
p = 2\mbox{pr}\left\{  Z > \max(|T_{lk}^+|, |T_{lk}^-|) \right\}, 
\end{equation}
where $Z \sim N(0, 1)$. 
A special scenario is to detect the presence/absence of correlations, i.e., $\xi = 0$.
In this case, $T^{+}_{lk} = T_{lk}^{-} = T_{lk} = \sqrt{n}\widehat{\delta}_{lk}^{-1}\widehat{r}_{lk}$, and the $p$-value in (\ref{test:stat2}) becomes the conventional two-sided $p$-value $p = 2\mbox{pr}\left\{Z > |T_{lk}| \right\}$.
Notably, like the direct estimation procedure, the proposed testing procedure can be directly implemented using lower-level data without performing any data aggregation.

The direct estimator, despite its advantages, encounters a notable limitation: it may not always be positive semi-definite (PSD), particularly when dealing with limited sample sizes. This issue arises because the direct estimation process is applied independently to each entry of the correlation matrix. Consequently, the resulting $\widehat{\Sigma}_{\text{dir}}$ does not conform to the form $M^\intercal M$ for some matrix $M$. This characteristic is in contrast to the classical sample covariance matrix, which inherently takes the $M^\intercal M$ form and is, therefore, always PSD as per basic linear algebra principles. The lack of guaranteed PSD in $\widehat{\Sigma}_{\text{dir}}$ poses challenges for downstream analyses that require the positive definiteness of $\widehat{R}_{\text{dir}}$, such as clustering analyses.
We address this limitation by developing a shrinkage procedure that shrinks $\widehat{\Sigma}_{\text{dir}}$ towards its diagonal entries $\text{diag}(\widehat{\Sigma}_{\text{dir}})$.
More specifically, we define
$\widehat{\Sigma}_{\text{dir-sh}} = \rho \text{diag}(\widehat{\Sigma}_{\text{dir}}) + (1 - \rho) \widehat{\Sigma}_{\text{dir}}$ for some $\rho \in (0, 1)$.
Let $\lambda_{\min}(M)$ denote the smallest negative eigenvalue of $M$ for any matrix $M$. 
Suppose $\lambda_{\min}(\widehat{R}_{\text{dir}}) < 0$. 
Since 
\(
\widehat{\Sigma}_{\text{dir-sh}} = \rho \text{diag}(\widehat{\Sigma}_{\text{dir}})^{1/2} \left\{ I_p + \rho^{-1}(1-\rho) \widehat{R}_{\text{dir}}\right\} \text{diag}(\widehat{\Sigma}_{\text{dir}})^{1/2}, 
\)
to ensure the positive definiteness of $\widehat{\Sigma}_{\text{dir-sh}}$, a sufficient condition is 
\(
\rho^{-1}(1-\rho)|\lambda_{\min}(\widehat{R}_{\text{dir}})| \leq 1/(1+\kappa)
\)
for some $\kappa > 0$, or equivalently, 
\begin{equation}\label{cond:1}
\rho \geq \frac{(1 + \kappa)|\lambda_{\min}(\widehat{R}_{\text{dir}})|}{1 + (1 + \kappa)|\lambda_{\min}(\widehat{R}_{\text{dir}})| }. 
\end{equation}
This leads to the following optimization problem: 
\begin{align}\label{opt:1}
\min_{\rho}\mathbb{E}\|\widehat{\Sigma}_{\text{dir-sh}} - \Sigma\|^2_{\text{F}}, \mbox{ s.t. (\ref{cond:1}) holds.} 
\end{align}
The following result provides the solution to (\ref{opt:1}). 
\begin{theorem}\label{opt:thm}
Suppose all assumptions in Theorem \ref{thm:cor} hold. 
Consider the scenario where $\lambda_{\min}(\widehat{R}_{\text{dir}}) < 0$. Then, for a pre-specified $\kappa > 0$, 
the solution to the optimization (\ref{opt:1}) is
$\widehat{\Sigma}_{\text{dir-sh}}(\kappa) = \rho_{\text{dir}}(\kappa)\text{diag}(\widehat{\Sigma}_{\text{dir}}) + (1 - \rho_{\text{dir}}(\kappa) )\widehat{\Sigma}_{\text{dir}}$, where
\begin{align}\label{opt:thm:eq1}
    \rho_{\text{dir}}(\kappa) = &~ \max\left\{ \frac{\gamma^2 - \beta^2}{\alpha^2 + \gamma^2 - 2\beta^2}, \frac{(1+\kappa)|\lambda_{\min}(\widehat{R}_{\text{dir}})|}{1 + (1+\kappa)|\lambda_{\min}(\widehat{R}_{\text{dir}})|} \right\}, 
\end{align}
where $\alpha^2 = \mathbb{E}\|\Sigma - \text{diag}(\widehat{\Sigma}_{\text{dir}})\|_{\text{F}}^2$,  
$\beta^2 = \sum_{l=1}^p \mathbb{E}(\widehat{\sigma}_{ll} - \sigma_{ll})^2$, and 
$\gamma^2 = \mathbb{E}\|\widehat{\Sigma}_{\text{dir}} - \Sigma\|_{\text{F}}^2$.
\end{theorem}
The proof is given in the supplementary document. 
By definition, it is not hard to see that $\alpha^2 > \beta^2$ and $\gamma^2 > \beta^2$, indicating that
$0 < \rho_{\text{dir}}(\kappa) < 1$.
One can see from the proof that, although $\rho_{\text{dir}}(\kappa) = (\gamma^2 - \beta^2)/(\alpha^2 - 2\beta^2 + \gamma^2)$ minimizes $\mathbb{E}\|\widehat{\Sigma}_{\text{dir-sh}} - \Sigma\|_{\text{F}}^2$, it may not ensure the positive definiteness of $\widehat{\Sigma}_{\text{dir-sh}}$. 
This indicates that our direct estimator $\widehat{\Sigma}_{\text{dir}}$ may need to be shrunk further towards its diagonal due to the potential negative eigenvalues. 

We next discuss estimation of $\alpha^2$, $\beta^2$, and $\gamma^2$.  
Some algebra yields 
\(
\alpha^2 = \|\Sigma\|_{\text{F}}^2 -2\mbox{tr}\left( \Sigma \text{diag}(\Sigma) \right) + \sum_{l=1}^p \sigma^2_{ll} + \beta^2.
\)
According to Lemma \ref{cov:upc:prop}, we have
\(
n \beta^2 \rightarrow \sum_{l=1}^p {|\mathcal{S}_{l}|^{-2}  (|\mathcal{S}_{l}| - 1)^{-2} } {\bf m}_{ll}^\intercal V {\bf m}_{ll}
\)
as $n \rightarrow \infty$. 
The above calculations lead to the following estimates
\[
\widehat{\beta}^2 = \sum_{l=1}^p \frac{1}{n|\mathcal{S}_{l}|^2  (|\mathcal{S}_{l}| - 1)^2} {\bf m}_{ll}^\intercal \widehat{V} {\bf m}_{ll}
\mbox{  and   } 
\widehat{\alpha}^2 = \|\widehat{\Sigma}_{\text{dir}}\|_{\text{F}}^2 -2\mbox{tr}\left( \widehat{\Sigma}_{\text{dir}} \text{diag}(\widehat{\Sigma}_{\text{dir}}) \right) + \sum_{l=1}^p \widehat{\sigma}^2_{ll} + \widehat{\beta}^2, 
\]
where $\widehat{V} = n^{-1}\sum_{i=1}^n {\bf z}_i{\bf z}_i^\intercal \otimes {\bf z}_i{\bf z}_i^\intercal - \mbox{vec}(\widehat{C}) \mbox{vec}(\widehat{C})^\intercal$ with $\widehat{C} = (\widehat{c}_{jk})$.
Similarly, since 
\[
{n\gamma^2} = n\sum_{l,k = 1, \ldots, p} \mathbb{E} \left\{ \widehat{\sigma}_{lk} - \sigma_{lk} \right\}^2 = n\sum_{l,k = 1, \ldots, p} \mbox{Var}\left( \widehat{\sigma}_{lk} \right)   \rightarrow \mbox{tr}(\Theta) \mbox{ as }  n \rightarrow \infty, 
\]
we 
then obtain $\widehat{\gamma}^2 = \mbox{tr}(\widehat{\Theta})/n$, where $\widehat{\Theta}$ is the estimate of $\Theta$ by replacing $V$ with $\widehat{V}$.  
Finally, we obtain the shrinkage correlation estimator
\begin{equation}\label{UPC}
\widehat{R}_{\text{dir-sh}}(\kappa) = \text{diag}\left(\widehat{\Sigma}_{\text{dir-sh}}(\kappa) \right)^{-1/2} \widehat{\Sigma}_{\text{dir-sh}}(\kappa) \text{diag}\left(\widehat{\Sigma}_{\text{dir-sh}}(\kappa)
\right)^{-1/2}. 
\end{equation}




In addition to ensuring positive definiteness, the shrinkage estimator has additional merits. 
First, when $\rho_{\text{dir}}(\kappa) = (\gamma^2 - \beta^2)/(\alpha^2 - 2\beta^2 + \gamma^2)$, one can verify that 
\begin{equation}\label{eq:shrink}
\frac{\mathbb{E}\|\widehat{\Sigma}_{\text{dir-sh}}(\kappa) - \Sigma \|_{\text{F}}^2}{\mathbb{E}\|\widehat{\Sigma}_{\text{dir}} - \Sigma \|_{\text{F}}^2} =  \frac{\alpha^2\gamma^2 - \beta^4}{\alpha^2\gamma^2 + \gamma^4 - 2\beta^2\gamma^2} < 1;
\end{equation}
here, we use the fact that $\gamma^2 > \beta^2$. 
Thus, the proposed shrinkage procedure is still appealing when $\widehat{\Sigma}_{\text{dir}}$ is already positive definite, because it provides a more accurate estimator of $\Sigma$ in terms of the expected squared Frobenius norm. 
Second, the entries of the shrinkage estimator $\widehat{R}_{\text{dir-sh}}(\kappa)$ keep the same relative magnitudes as the original direct estimator $\widehat{R}_{\text{dir}}$ for any $\kappa > 0$.  
To see this, consider two pairs of proteins $(l_1, k_1)$ and $(l_2, k_2)$, and
some calculations yield 
\[
\frac{\left( \widehat{R}_{\text{dir-sh}} \right)_{l_1k_1}}{\left( \widehat{R}_{\text{dir-sh}} \right)_{l_2k_2}} = \frac{ (1-\rho_{\text{dir}}(\kappa)) \widehat{\sigma}_{l_1k_1} }{\sqrt{\widehat{\sigma}_{l_1l_1}\widehat{\sigma}_{k_1k_1}} }
\times \frac{\sqrt{\widehat{\sigma}_{l_2l_2}\widehat{\sigma}_{k_2k_2}} }{{ (1-\rho_{\text{dir}}(\kappa)) \widehat{\sigma}_{l_2k_2} }} = \frac{\widehat{r}_{l_1k_1}}{\widehat{r}_{l_2k_2}}. 
\]

Finally, we discuss how we choose $\kappa$. 
We propose a cross-validation method analogous to the one used in \cite{bickel2008covariance}. 
We split the sample randomly into two pieces of size $n_1$ and $n_2$. 
Let $\widehat{\Sigma}^{(b)}_{1, \text{dir}}$ and $\widehat{\Sigma}^{(b)}_{2, \text{dir}}$ be the proposed direct estimator of $\Sigma$ based on the $n_1$ and $n_2$ observations respectively from the $b$-th split. 
We define 
\[
\widehat{\text{CV}}(\kappa) = \frac{1}{B} \sum_{b=1}^B \| \widehat{\Sigma}^{(b)}_{1, \text{dir-sh}} (\kappa) -  \widehat{\Sigma}^{(b)}_{2, \text{dir}}\|_F^2,
\]
where $\widehat{\Sigma}^{(b)}_{1, \text{dir-sh}} (\kappa)$ is the shrinkage estimator based on $\widehat{\Sigma}^{(b)}_{1, \text{dir}}$. 
Like other cross-validation procedures, we consider a pool of $\kappa$ values and choose the optimal $\kappa$ that minimizes $\widehat{\text{CV}}(\kappa)$.

\section{Simulation studies}\label{sec:simu}

In this section, we evaluate the performance of our direct estimation procedures (denoted as DIR and DIR$\_\text{sh}$ for $\widehat{R}_{\text{dir}}$ and $\widehat{R}_{\text{dir-sh}}(\kappa)$, respectively) in terms of estimation accuracy. 
{ For  $\widehat{R}_{\text{dir-sh}}(\kappa)$, $\kappa$ was determined by the proposed cross-validation procedure.} 
We also assessed the finite-sample performance of the direct inference procedure, focusing on the type-I error rate and power.
We compare our proposed method with several data-aggregation-based methods, including the Sum of Unique Variables (SUV), Mean of Unique Variables (MUV), Sum of All Variables (SAV), Mean of All Variables (MAV), Tukey Median Polish using either all variables (TMP-all) or only unique variables (TMP-uni), Singular Value Decomposition with all variables (SVD-all) or unique variables (SVD-uni), Sum of the Three most Intense unique variables (STI), Mean of the Top 50\% of all variables (MT50), and SCAMPI.
For each aggregation method, we initially computed the estimates for the higher-level variables and then calculated the sample covariance or correlation matrix. From these estimated correlations, we derived $p$-values to test the hypothesis outlined in \eqref{hypo:1} using the Fisher transformation, implemented via the {\tt cor.test()} function in R.


In our simulation study, we first generated the true covariance matrix $\Sigma$ for higher-level variables, considering two scenarios with $p=20$ and $50$ variables, respectively. 
The structure of $\Sigma$ was based on a network where $70\%$ of the off-diagonal entries were non-zero, each generated from a uniform distribution over the interval $[0.2, 0.5]$. To ensure positive definiteness, we set the diagonal entries of $\Sigma$ to 1.5 for $p=20$ and to 2.5 for $p=50$. The covariance matrix was then normalized to obtain the correlation matrix $R$ using the transformation $\mbox{diag}(\Sigma)^{-1/2}\Sigma \mbox{diag}(\Sigma)^{-1/2}$. The maximum correlations between higher-level variables in these simulations were 0.33 for $p=20$ and 0.19 for $p=50$.
For the link matrix $A \in \mathbb{R}^{q \times p}$, we simulated $q$ lower-level variables ($q > 5p$), ensuring each higher-level variable was associated with 5 unique lower-level variables. The sizes of $q$ were set to 150 for $p=20$ and 300 for $p=50$. We conducted 500 independent replications for each setting. In each replication, the lower-level data ${\bf z}_i \in \mathbb{R}^{q}$ was generated from a multivariate normal distribution with mean ${\bf 0}$ and covariance $C = A\Sigma A^\intercal + \Gamma$, where $\Gamma$ is a diagonal matrix with all entries set to $0.3$.

\begin{figure}[!t]
   \centering
  \includegraphics[width=0.7\textwidth]{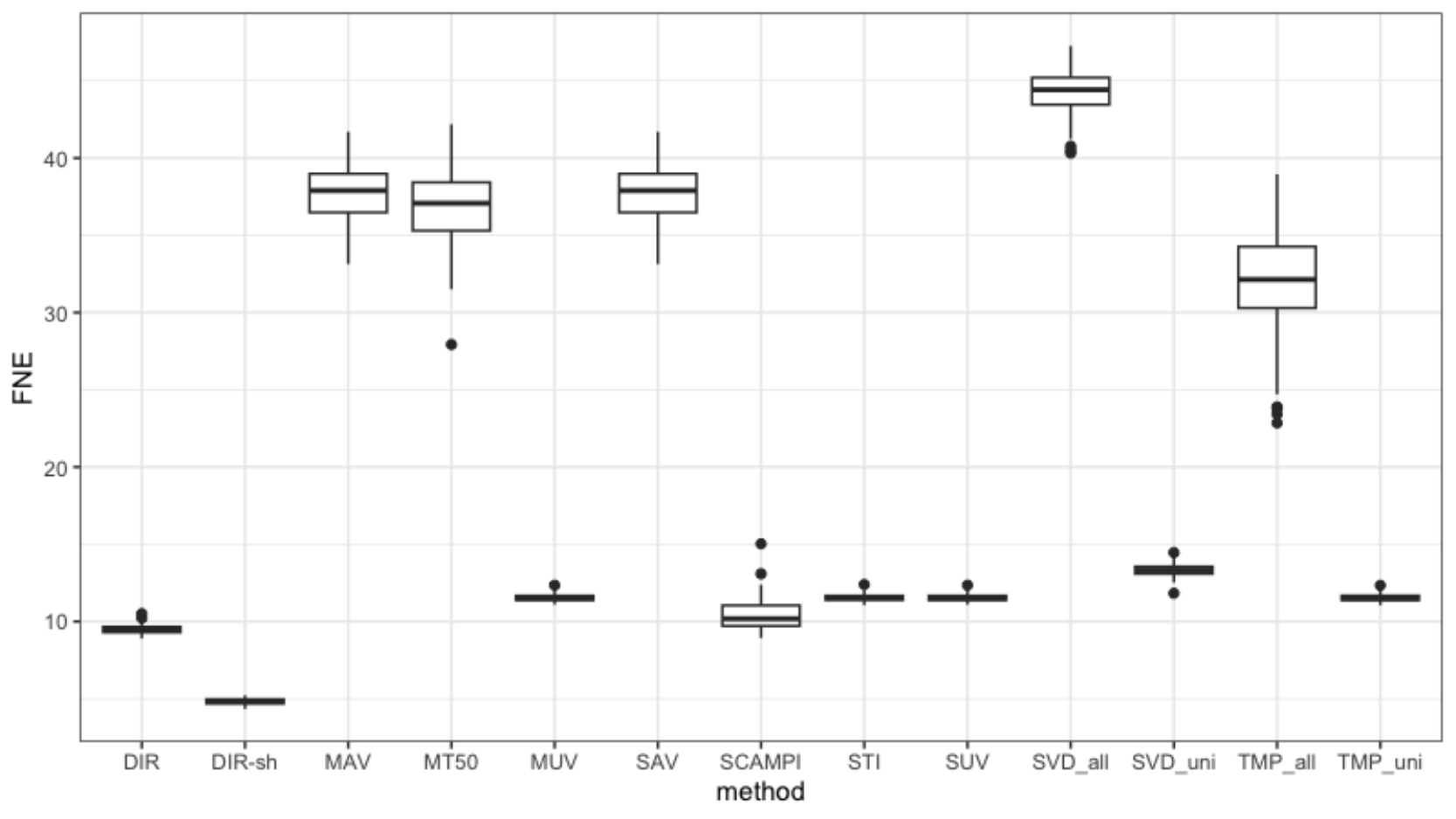}
  \caption{Boxplots of the FNE values over 500 replications with $n = 30$ and $p=50$.
  The proposed direct methods outperform existing aggregation-based methods.}
  \label{fig:simu3}
\end{figure}

Our first goal is to examine the estimation accuracy of all methods
 in terms of the Frobenius-norm error (FNE) according to
\(
\mbox{FNE}(\widehat{R}) = \|\widehat{R} - R\|_F
\)
for any estimator $\widehat{R}$. 
{ Since different values of $\kappa$ in $\widehat{R}_{\text{dir-sh}}(\kappa)$ only make a difference when $\lambda_{\min}(\widehat{R}_{\text{dir}}) < 0$, we considered a ``small $n$, large $p$" scenario with $p=50$ and $n=30$.}
From Fig. \ref{fig:simu3}, 
we can see that the direct method (both $\widehat{R}_{\text{dir}}$ and $\widehat{R}_{\text{dir-sh}}$) consistently outperform all aggregation-based methods, with the shrinkage estimator showing the highest estimation accuracy. 
{ This demonstrates the effectiveness of the cross-validation procedure for determining $\kappa$. }
Secondly, among the existing methods, SCAMPI emerges as the best performer. Thirdly, methods based on unique variables (MUV, SUV, TMP\_uni, SVD\_uni, STI) generally perform better than those utilizing all variables.
 More comparisons with existing latent factor models \citep{bai2012statistical, fan2016projected} and unsupervised GSEA methods \citep{barbie2009systematic, hanzelmann2013gsva, tomfohr2005pathway} can be found in the supplement. 



\begin{figure}[!t]
   \centering
  \includegraphics[ trim=2.5cm 0cm 4cm 0cm, clip, width=0.9\textwidth]{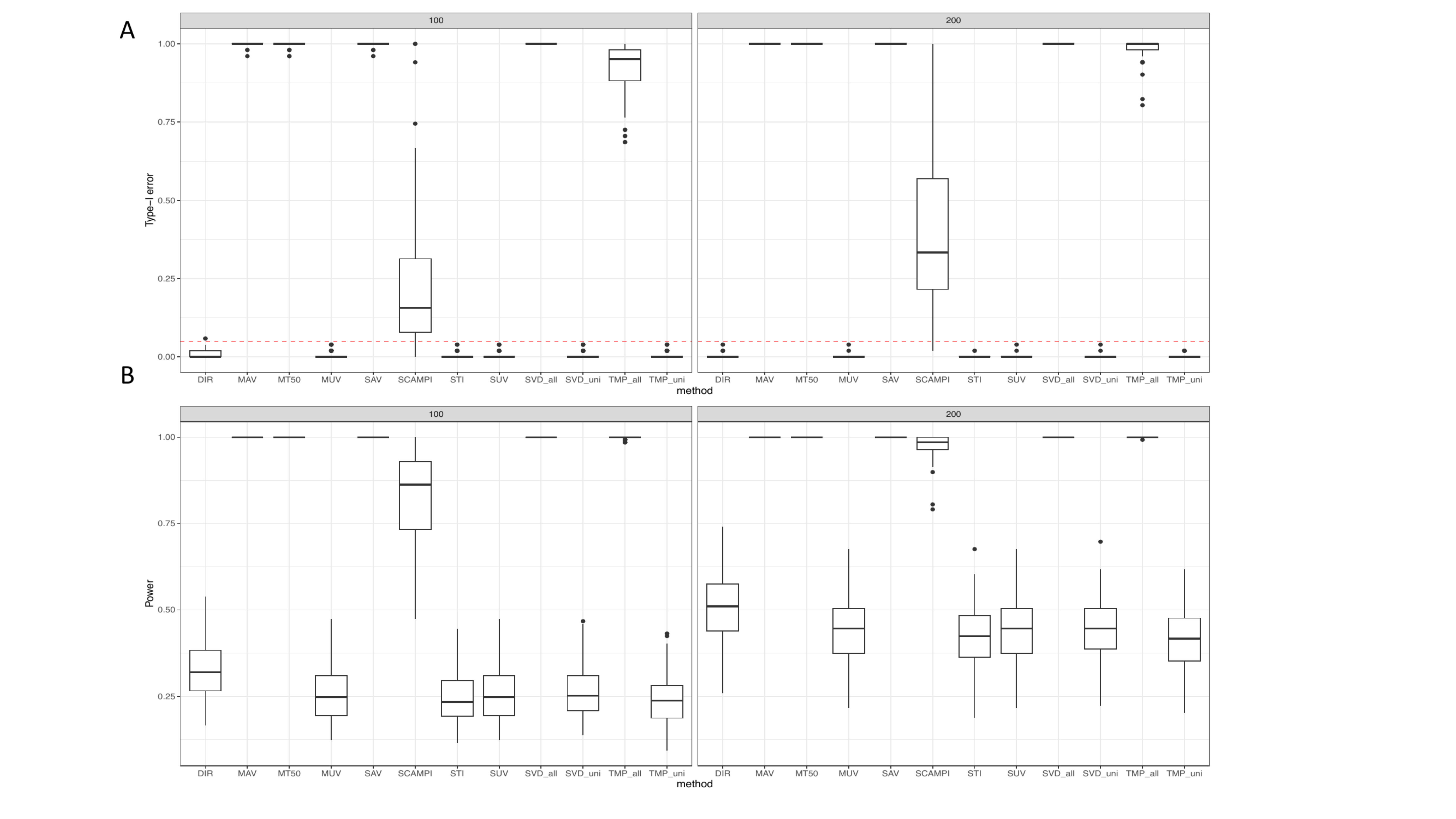}
  \caption{Boxplots of the type-I error and power over 500 replications with $n = 100, 200$: (A) type-I error with $p = 20$, (B) power with $p = 20$. The red dashed line in Fig. \ref{fig:simu2}A is at 0.05.
  Aggregation methods using all variables have type-I error rates close to 1. SCAMPI has inflated type-I errors.
  The proposed direct inference has controlled type-I error rates and has higher power than other methods that also control type-I error rates.}
  \label{fig:simu2}
\end{figure}

Our second objective was to test the hypotheses $H_{0, lk}: |r_{lk}| \leq 0.1$ against $H_{1, lk}: |r_{lk}| > 0.1$ for $l = 1, \ldots, p-1$ and $k = l+1, \ldots, p$. 
All tests were conducted at a two-sided significance level of $\alpha = 0.05$.
The results for sample sizes $n = 100$ and 200 and $p=20$ are presented in Fig. \ref{fig:simu2}. Fig. \ref{fig:simu2}A reveals that our proposed direct method, along with unique-variable-based methods (MUV, SUV, TMP\_uni, SVD\_uni), maintains well-controlled type-I error rates. Conversely, other methods, particularly MAV, SAV, and MT50, exhibit significantly inflated type-I errors, often approaching 1. Fig. \ref{fig:simu2}B highlights that among the methods with controlled type-I error rates, the direct method demonstrates the highest power.
It's noteworthy that methods with controlled type-I error rates generally exhibit relatively low power, as depicted in Fig. \ref{fig:simu2}B. This is attributed to the inherently weak signals in the dataset, with the maximum correlation capped at 0.33 for $p = 20$, indicating that detecting such subtle correlations may require larger sample sizes. 
Comparisons with ssGSEA, GSVA, and PLAGE
in ``more extreme" scenarios, including settings with more overlapping/shared variables, smaller sample sizes, and potentially confounding variables, can be found in the supplement. 

\section{Real Data Application}\label{sec:realdata}

\subsection{Gene-pathway application}\label{subsec:kegg}
In this application, our goal is to estimate the correlations between gene pathways differentially expressed across various stages of lung cancer. We utilized a gene expression dataset from The Cancer Genome Atlas (TCGA) Program \citep{cancer2014comprehensive}, specifically focusing on lung cancer patients. 
This dataset, sourced from the Lung Cancer Explorer (LCE) with standard quality control, includes log 2-transformed expression levels of 20,429 genes from 513 lung cancer tissue samples. The sample distribution across cancer stages is as follows: 278 patients with stage I, 124 with stage II, 84 with stage III, and 27 with stage IV.
We considered 127 common human pathways from the KEGG database \citep{kanehisa2000kegg}, which were detected by \cite{jin2022t2} as differentially expressed pathways between Stage I and Stage III using the T2-DAG $\chi^2$ test. 
These pathways encompass 2,787 genes, comprising 1,633 unique genes and 1,142 shared genes, with each pathway including 17 to 266 genes.
Of the 127 pathways, 109 meet UVC, meaning they contain at least two unique genes. The number of unique genes in each of these 109 pathways varies from 2 to 91. It's noteworthy that none of the 18 pathways failing to meet UVC are among the top 30 most significant pathways identified in the analysis.  

For this application, we estimated the correlations between 109 pathways using our proposed direct method, analyzing Stage-I (278 patients) and Stage-III (84 patients) lung cancer separately. In addition, we evaluated existing aggregation-based methods, excluding STI due to the lack of three unique genes in all gene pathways. Based on our simulation study findings, we implemented only MUV and MAV, as SUV and SAV yield equivalent correlation estimates.
 Although ssGSEA, GSVA, or PLAGE were not designed for correlation estimation between gene pathways, we also added them into the comparison. 
We tested the null hypothesis for different values of $\xi$ (ranging from 0.1 to 0.5) and reported the percentages of significant pathway pairs in Table \ref{table:KEGG_percent:grp1}.

Consistent with our simulation results, methods using all genes (MAV, TMP-all, SVD-all) and 
GSEA methods (ssGSEA, GSVA, PLAGE) 
tended to identify more significant correlations, likely resulting in false positives. 
For Stage-I patients, MUV detected more pathway pairs than the direct method at $\xi = 0.1$, possibly because MUV's correlations, while weaker than actual (see Proposition S1 in the supplementary document), remain significant at lower $\xi$ values. However, as $\xi$ increases, methods using only unique genes, including MUV, become more conservative compared to the direct method, indicating reduced power due to neglect of shared genes.
SCAMPI identified more significant correlations than the direct method at $\xi = 0.1$, but as indicated in Fig. \ref{fig:simu2} and Fig. S2 in the supplementary document, these may include false positives. Furthermore, SCAMPI is less powerful than the direct method in detecting strongly correlated pathways for $\xi > 0.1$. 
Similar patterns were observed for Stage-III patients.
Regarding computational efficiency, most existing aggregation methods completed within 5 seconds, relying on simple mathematical operations (e.g., sum, mean) and the efficient {\tt R} function {\tt cor.test()} for $p$-value calculation. SCAMPI, implemented in the {\tt R} package {\tt protiq}, required approximately 23 hours to estimate gene pathway expression levels for a single patient. In contrast, our direct method completed its estimation in about 10 seconds, with $p$-value calculations taking around 30 minutes. This longer duration for $p$-value calculations is primarily due to the computation of a large matrix $\widehat{V} \in \mathbb{R}^{q^2 \times q^2}$ (where $q=2,787$). We utilized efficient algorithms and packages for large-matrix computation to mitigate potential memory issues.

\begin{figure}
    \includegraphics[trim=1cm 2.5cm 1cm 2cm, clip, width=0.9\textwidth]{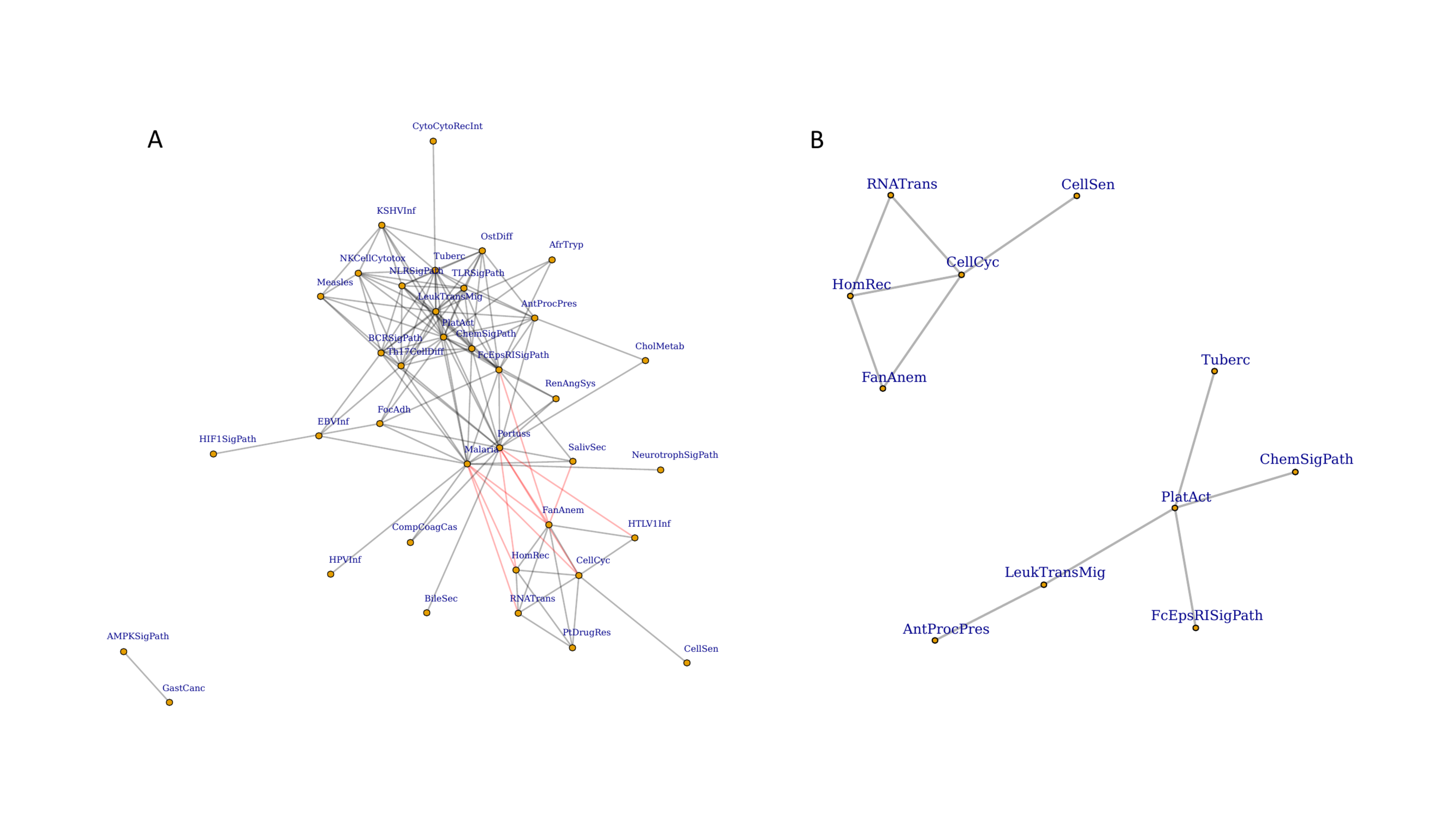}
    \caption{Network visualizations of the gene-pathway correlations stronger than 0.5: A: Stage I; B: Stage III.  
    Positive correlation values are shown in grey, while negative values are shown in red. 
    Abbreviated pathway names are used, and their corresponding full pathway names are given in Supplementary Tables S1 and S2. 
    }
    \label{fig:realdata2}
\end{figure}

To delve deeper into the biological significance of the highly correlated gene pathways in lung cancer patients, we visualized co-expression networks of significantly correlated pathways derived from the direct method with $\xi=0.5$. These are depicted in Fig. \ref{fig:realdata2}A and B for Stage-I and Stage-III patients, respectively. It's notable that the network for Stage-III patients shows fewer edges compared to Stage-I, which may be attributed to the sample size discrepancy (287 Stage-I patients vs. 84 Stage-III patients) rather than a decrease in correlation intensity at later cancer stages.
In Stage-I patients (Fig. \ref{fig:realdata2}A), several pathways, including PlactAct (Platelet Activation), LeukTransMig (Leukocyte Transendothelial Migration), ChemSigPath (Chemical Carcinogenesis - Reactive Oxygen Species Pathway), Tuberc (Tuberculosis), Pertuss (Pertussis), Malaria, FanAnem (Fanconi Anemia Pathway), and CellCyc (Cell Cycle), emerge as hubs. Notably, Pertuss, Malaria, FanAnem, and CellCyc exhibit negative correlations with various other pathways. Many of these hub pathways, specifically PlactAct, LeukTransMig, ChemSigPath, Tuberc, FanAnem, and CellCyc, are also prominent in Stage-III patients (Fig. \ref{fig:realdata2}B), with PlactAct and CellCyc playing central roles. However, unlike in Stage-I, only positive correlations are observed in Stage-III.
These findings align with existing research connecting these pathways to cancer progression. 
For example, 
PlatAct contributes to the initiation of events promoting tumorigenesis, such as the release of growth factors and angiogenic factors, influencing immune, stromal, and tumor cells \citep{bambace2011platelet}. 
Tuberc has been considered an important risk factor for lung cancer \citep{qin2022relationship}.
Mutations in FanAnem have also been suggested to link to cancer susceptibility \citep{liu2020fanconi}. 

\begin{table}[ht]
\centering
\caption{The percentages (\%) of significant pairs of gene pathways for $\xi =  0.1, 0.2, \ldots, 0.5$ in Stage-I ($n=278$) and Stage-III ($n=84$) patients: 
All percentages were rounded to one decimal place.}
\label{table:KEGG_percent:grp1}
\begin{tabular}{|c|cc|cc|cc|cc|cc|}
\hline
Method & \multicolumn{2}{|c|}{$\xi = 0.1$} & \multicolumn{2}{|c|}{$\xi = 0.2$} & \multicolumn{2}{|c|}{$\xi = 0.3$} & \multicolumn{2}{|c|}{$\xi = 0.4$} & \multicolumn{2}{|c|}{$\xi = 0.5$} \\
\hline
 & I & III & I & III & I & III & I & III & I & III \\
\hline
Direct & 31.2 & 12.5 & 17.4 & 5.9 & 10.3 & 3.1 & 6.5 & 1.6 & 4.2 & 1.0 \\
 MUV &  31.4 & 14.5 & 16.2 & 6.4 & 8.1 & 2.5 & 3.6 & 1.1 & 1.8 & 0.4\\
 MAV &   66.9 &  48.9 &  52.9 & 36.5 & 40.9  & 27.1 & 30.8 & 18.6 & 22.0 & 11.8\\
 TMP-all &   67.2 & 52.6 &   54.1 & 40.7 & 42.0 & 29.5 & 31.1 & 20.1 & 21.6 & 12.1\\
 TMP-uni & 26.4 & 12.1 &  13.2 & 5.2 & 6.8 & 2.4 & 2.9 & 0.9 & 1.5 & 0.4\\
 SVD-all &    72.5 & 55.9 &  59.2 & 41.8 & 45.4 & 28.2 & 32.6 & 17.4 & 20.1 & 8.9\\
 SVD-uni &   26.8 & 13.9 &  12.4 & 5.1 & 5.7 & 1.9 & 2.1 & 0.8 & 0.6 & 0.3\\
 MT50 &  64.1 & 47.8 &  50.1 & 35.5 & 37.6 & 24.4 & 25.6 & 14.6 & 15.8 & 6.9\\
 SCAMPI & 35.7 & 15.8 & 17.4 & 5.9 & 7.7 & 1.7 & 3.2 & 0.6 & 1.1 & 0.2\\
 GSVA & 55.7 & 38.7 & 39.8 & 24.8 & 26.4  & 14.4 & 15.5 & 7.6 & 8.1 & 3.7 
 \\ 
 PLAGE & 78.5 & 60.4 & 68.0  & 50.7 & 58.2 & 41.9 & 48.8 & 34.2 & 39.1 & 26.6\\
 ssGSEA & 59.2 & 40.5 & 44.3 & 26.0 & 30.9 & 14.7 & 18.9 & 8.2 & 10.8 & 4.0\\
 
\hline
\end{tabular}
\end{table}

\subsection{Proteomics application}\label{subsec:prot}

In this application, we consider a brain tumor proteomics data set from 227 subjects.
 The original data contains 5433 peptides belonging to 1803 protein groups (genes).
Among the 1803 genes, we focused on 175 genes found in the 206 common human pathways related to signaling, metabolic and human diseases from the KEGG database \citep{kanehisa2017kegg}. 
As a result, 1031 peptides that mapped into 175 protein groups were retained in the analysis.
All protein groups have multiple unique peptides (range: 2-34; median: 4), 
and 56 peptides are shared across different protein groups. 
To mitigate data skewness and correct for experimental variation, we applied log-transformation followed by median normalization to the peptide intensities \citep{ting2009normalization}.
The objective here is to identify highly correlated protein pairs relevant to brain cancer pathology. We compared our direct method with existing aggregation-based methods, mirroring the analysis approach outlined in Section \ref{subsec:kegg}. The findings are presented in the left part of Table \ref{table:PPI_percent}. 
Our direct approach detected more significant protein pairs than all aggregation-based methods utilizing unique peptides. However, unlike the results in Tables \ref{table:KEGG_percent:grp1}, MAV, TMP-all, and SVD-all only identified slightly more protein pairs than their unique-peptide counterparts (MUV, TMP-uni, and SVD-uni). 
This might be attributed to the lesser impact of shared peptides due to their lower count (56) in this dataset compared to the 1142 shared genes in the KEGG pathways dataset.
Differing from our simulations and the KEGG application, SCAMPI exhibited the lowest power among all methods, suggesting variability in its performance across different applications. 
To investigate potential false positives detected by existing methods, we evaluated the overlap between protein pairs detected by these methods and those identified by our direct method. This comparison, focusing on the percentage of common detections, is presented in the right part of Table \ref{table:PPI_percent}. 
The unique-peptide-based methods (MUV, TMP-uni, SVD-uni) show high consistency with the direct method, particularly at higher values of $\xi$. In contrast, methods using all peptides (MAV, TMP-all, SVD-all, MT50) demonstrate lower consistency, indicating potential false positives. Interestingly, while SCAMPI's consistency with the direct method is lowest at $\xi=0.1$, it improves with increasing $\xi$ values.
In terms of computational efficiency, all existing aggregation methods, excluding SCAMPI, completed within 2 seconds. SCAMPI, in this instance, took about 6 minutes per patient for estimating protein abundances, a marked improvement from the previous application. This speed suggests SCAMPI may be more suitable for datasets with relatively smaller dimensions of lower-level variables. Our direct method completed its estimation in 2 seconds and required approximately 5 minutes for $p$-value calculation, reflecting its efficiency in handling the 1075 peptides in this dataset.

\begin{figure}
    \includegraphics[trim=2cm 3cm 2cm 2cm, clip, width=0.3\textwidth]{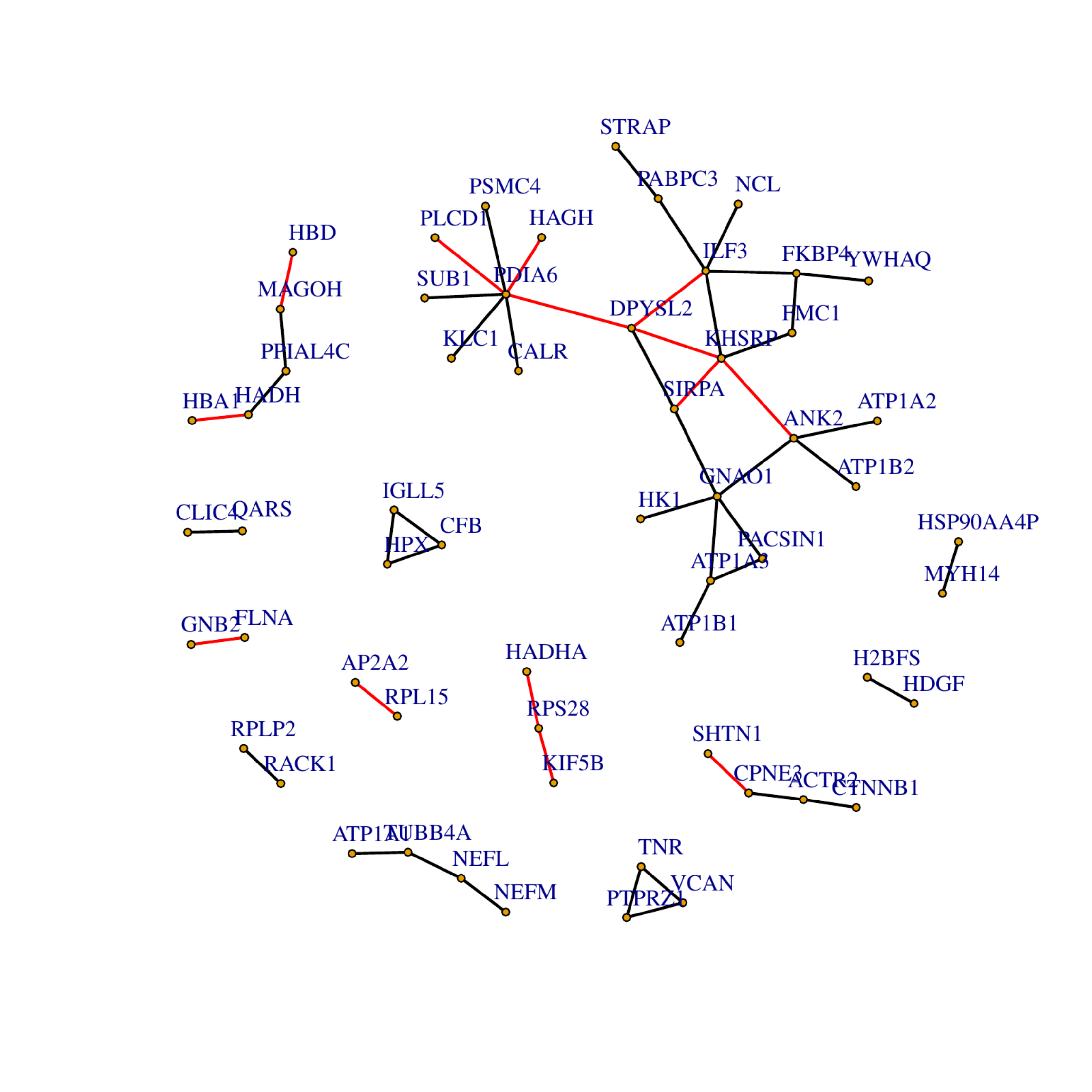}
    \centering
    \caption{A network visualization of the protein correlations stronger than 0.5. 
    The layout of the nodes is generated by {\tt igraph} for optimal visualization \citep{csardi2006igraph}. Positive correlation values are shown in black, while negative values are shown in red.    }
    \label{fig:realdata1}
\end{figure}

\begin{table}[ht]
\centering
\caption{On the left, percentages (\%) of significant protein pairs identified by all methods were presented. 
On the right, percentages (\%) of significant protein pairs that were also detected by the direct method for each aggregation-based method: $\xi = 0.1, 0.2, \ldots, 0.5$;
all percentages were rounded to one decimal place.}
\label{table:PPI_percent}
\begin{tabular}{|c|ccccc|ccccc|}
\hline
Method & {$\xi = 0.1$} & {$\xi = 0.2$} & {$\xi = 0.3$} & {$\xi = 0.4$} & {$\xi = 0.5$} & \multicolumn{5}{|c|}{ Comparison } \\
\hline
Direct &  25.2 & 14.8 & 8.0  & 4.1 &  2.1 & &  &  &  & \\
 MUV &   20.4 & 7.1 & 2.3 & 0.6  & 0.1 &  85.1 & 87.8 & 90.4 & 92.5 & 100\\
 MAV &   21.2  & 7.7  & 2.7 & 0.9  & 0.2 &  79.3 & 78.7 & 77.3 & 70.9 & 67.6 \\
 TMP-all  & 21.1  & 7.7  & 2.6  & 0.8  & 0.3 &  66.9 & 69.8 & 70.1 & 62.7 & 54.5\\
 TMP-uni  & 20.1  & 7.1  & 2.1  & 0.5 & 0.1 & 71.2 & 76.0 & 82.4 & 85.7 & 90.9\\
 SVD-all  & 21.3  & 7.8 & 2.6  & 0.9  & 0.2 &78.6 & 77.4 & 75.9 & 69.3 & 63.4\\
 SVD-uni & 20.4  & 7.2  & 2.2  & 0.6  & 0.1 & 84.7 & 87.1 & 90.4 & 93.6 & 100 \\
 MT50 & 22.5  & 8.4  & 2.5 & 0.8 & 0.3 & 57.1 & 57.4 & 59.8 & 48.8 & 39.0\\
 SCAMPI  & 12.2  & 4.4 & 1.3  & 0.3 & $<$0.1 & 66.8 & 70.9 & 76.1 & 76.7 & 88.9\\
\hline
\end{tabular}
\end{table}
To derive biological insights from the protein correlations identified, we visualized a protein-protein interaction network using the direct method for $\xi=0.5$, as shown in Fig. \ref{fig:realdata1}. This network represents proteins by their gene identifiers, with edges indicating significant protein correlations. Positive correlations are marked in black, while negative correlations are in red. Isolated nodes were excluded from the visualization.
The network in Fig. \ref{fig:realdata1} highlights several key hub nodes, including PDIA6, ILF3, KHSRP, and GNAO1. PDIA6, known for its role in protein folding, is crucial for maintaining cellular homeostasis and has been linked to poor outcomes in various cancers \citep{ma2021pdia6}. ILF3, a double-stranded RNA-binding protein, is significant in immune response regulation and is being explored as a prognostic marker in different cancers \citep{liu2019ilf3}. KHSRP, another RNA-binding protein, plays an influential role in immune cell function and tumor progression \citep{palzer2022role}. Lastly, GNAO1, which is highly expressed in the mammalian brain, is involved in the transduction of G-protein-coupled receptor (GPCR) signals, underscoring its potential relevance in neurological pathologies, including brain cancers \citep{ling2022treating}.

\section{Discussions}
This paper introduces a novel estimation and inference procedure for directly investigating higher-level correlations based on lower-level measurements.
Under the UVC assumption, our latent factor model can correctly capture the relationship between lower-level and higher-level variables. 
As such, the proposed direct estimator has superior performance by efficiently leveraging information from all unique and shared variables. 
Furthermore, we have developed a shrinkage procedure that yields a positive-definite correlation estimator, optimizing estimation accuracy.
  
Looking ahead, our proposed latent factor model can be further combined with an additional outcome model to investigate the association between the higher-level variables and an outcome of interest. 
Specifically, consider
\[
{\bf z}_i = A{\bf x}_i + \ve \varepsilon_i, y_i = {\bf x}_i^\intercal \ve \beta + \delta_i,
\]
where $y_i$ is the disease outcome, and $\ve \beta$ is the parameter of interest. 
Some calculations yield that $\ve \beta = \Sigma^{-1}\mathbb{E}[y_i{\bf x}_i]$ and $\mathbb{E}[y_i{\bf x}_i] = (A^\intercal A)^{-1} \mathbb{E}[y_i {\bf z}_i]$. 
Therefore, a direct estimator of $\ve \beta$ is
\(
\widehat{\ve \beta} = \widehat{\Sigma}_{\text{dir-sh}}^{-1} (A^\intercal A)^{-1} \{n^{-1} \sum_{i=1}^n y_i {\bf z}_i\}.
\)
Notably, computing $\widehat{\ve \beta}$ does not necessitate the quantification of the higher-level variables ${\bf x}_i$. Theoretical and numerical analyses of $\widehat{\ve \beta}$ are avenues for our future research.



\section{Supplementary Material}
\label{sec6}

Supplementary material is available online at
\url{http://biostatistics.oxfordjournals.org}. 
The {\tt R} package {\tt highcor} for implementing the proposed direct estimation and inference methods is available for download at \url{https://github.com/taryue/highcor}.

\bibliographystyle{chicago}
\bibliography{reference, refs}

\end{document}